\documentclass[prl,twocolumn,showpacs,preprintnumbers,amsmath,amssymb]{revtex4}

\usepackage{graphicx}
\usepackage{dcolumn}
\usepackage{bm}
\usepackage{amsmath,amssymb}
\usepackage{mathrsfs}
\usepackage{color}

\begin{document}

\title{Photon-Photon Entanglement with a Single Trapped Atom}

\author{B.~Weber}
\author{H.~P.~Specht}
\author{T.~M{\"{u}}ller}
\author{J.~Bochmann}
\author{M.~M{\"{u}}cke}
\author{D.~L.~Moehring}
 \email{david.moehring@mpq.mpg.de}
\author{G.~Rempe}

\affiliation{Max-Planck-Institut f{\"{u}}r Quantenoptik, Hans-Kopfermann-Strasse~1, 85748 Garching, Germany}

\date{\today}

\begin{abstract}
An experiment is performed where a single rubidium atom trapped within a high-finesse optical cavity emits two independently triggered entangled photons.  The entanglement is mediated by the atom and is characterized both by a Bell inequality violation of $S=2.5$, as well as full quantum-state tomography, resulting in a fidelity exceeding $F=90\%$.  The combination of cavity-QED and trapped atom techniques makes our protocol inherently deterministic --- an essential step for the generation of scalable entanglement between the nodes of a distributed quantum network. 
\end{abstract}


\maketitle

Of all the technologies currently being pursued for quantum information science, individually trapped atoms are among the most proven candidates for quantum information storage \cite{monroe:2002}.  Photons, on the other hand, are the obvious choice for carriers of quantum information over large distances.  Together, this naturally leads to an atom-photon interface as an ideal node for distributed quantum computing networks \cite{cirac:1997, monroe:2002, kimble:2008}.  Progress towards the construction of such quantum networks has been recently achieved in experiments entangling single atoms trapped in a free-space radiation environment with their spontaneously emitted photons \cite{blinov:2004, volz:2006, matsukevich:2008, rosenfeld:2008}, however, high photon loss rates in the emission process severely limit their usefulness for quantum information processing protocols \cite{campbell:2007}.  For \textit{scalable} atom-photon based quantum information processing, it is necessary to increase this entanglement efficiency.  The most promising method to accomplish this is to combine the advantages of trapped atom entanglement techniques with cavity quantum electrodynamics where both atomic and photonic qubits are under complete control \cite{kimble:2008, wilk:2007b, hijlkema:2007, fortier:2007, khudaverdyan:2008}. 

In this Letter, we demonstrate a deterministic entanglement protocol with a single atom trapped in an optical cavity and two subsequently emitted single photons.  Compared to previous entanglement experiments with a probabilistic transit of atoms through a cavity \cite{wilk:2007b}, our results increase the atom-cavity interaction time, and therefore also the number of successful atom-photon entanglement events from a single atom, by a factor of $10^5$.  The long trapping times shown here also allow us to ensure that exactly one atom is within the cavity at a given time.  This is critical for the generation of high-fidelity entangled states, and is not possible with atoms randomly loaded into a cavity \cite{wilk:2007b}.  Furthermore, the highly efficient photon collection in the cavity output mode allows for photon detection efficiencies that are more than an order of magnitude greater than in free-space atom-photon entanglement experiments \cite{matsukevich:2008, rosenfeld:2008}.  This also allows for the coherent mapping of the atomic quantum state onto the state of a second photon.  The resulting entanglement is verified by a Bell inequality measurement between the two emitted photons \cite{bell:1964}, and is in convincing violation of classical physics.  

\begin{figure}
\includegraphics[width=1\columnwidth,keepaspectratio]{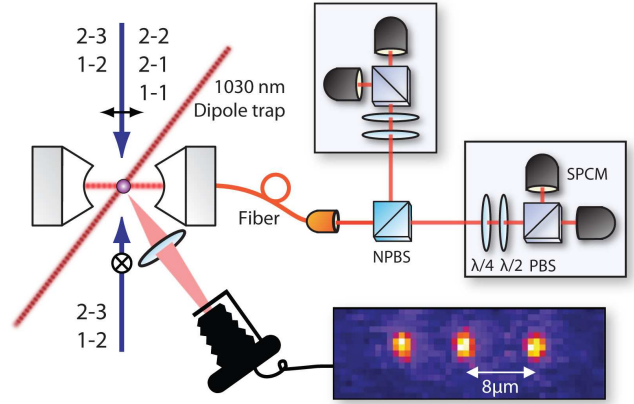}
\caption{\label{fig:setup} Individual $^{87}$Rb atoms are trapped within the TEM$_{00}$ mode of a high-finesse optical cavity (finesse $\approx3\times10^4$) at the intersection of two standing-wave dipole trap beams. Lin$\bot$lin-polarized laser beams orthogonal to the cavity axis provide motional cooling, while additional beams for optical pumping and the creation of entangled photons are polarized along the cavity axis and independently directed onto the atom.  The cavity output is coupled into an optical fiber and directed to the photonic state detection apparatus.  Perpendicular to the cavity axis, a CCD camera is used to monitor the atoms within the trap.  The displayed image shows three atoms coupled to the mode of the cavity and aligned along the 1030~nm beam.  SPCM: single photon counting module, NPBS: non-polarizing beam splitter, PBS: polarizing beam splitter, $\lambda/4$: quarter-wave plate, $\lambda/2$: half-wave plate.}
\end{figure}

\begin{figure*}
\includegraphics[width=2.0\columnwidth,keepaspectratio]{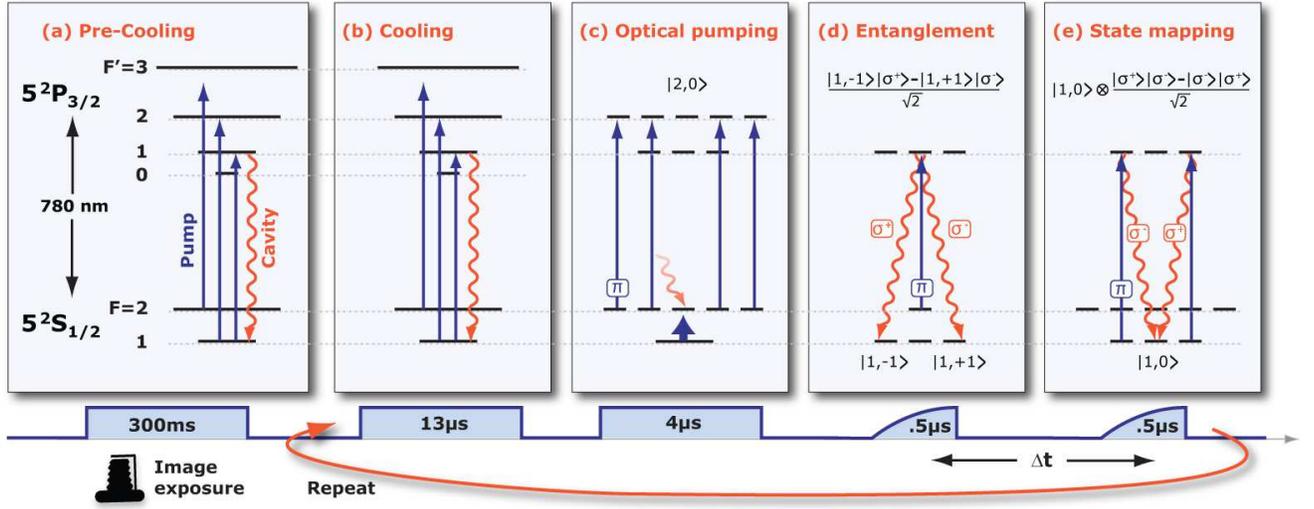}
\caption{\label{fig:pulses} 
The experimental procedure.
(a) When atoms are first loaded into the cavity, a 300~ms laser pulse is applied for optical cooling.  During this time, the cavity mode is simultaneously imaged with a camera to confirm the presence of a single atom.  
(b-e) The entanglement generation protocol runs at a repetition rate of 50~kHz.
(b) Atomic re-cooling.
(c) A $\pi$-polarized laser resonant with the $F$$=$$2\leftrightarrow$~$F'$$=$$2$ laser together with resonant lasers on the $F$$=$$1\leftrightarrow$~$F'$$=$$1$ and $F$$=$$1\leftrightarrow$~$F'$$=$$2$ transitions optically pump the atom to the $|F$$=$$2,m_{F}$$=$$0\rangle$ Zeeman sublevel.
(d) A $\pi$-polarized $F$$=$$2\leftrightarrow$~$F'$$=$$1$ laser transition generates atom-photon entanglement.
(e) After a time $\Delta t$, a $\pi$-polarized $F$$=$$1\leftrightarrow$~$F'$$=$$1$ laser subsequently maps the quantum state of the atom onto a second photon.
}
\end{figure*}

The main element of our experimental apparatus is a coupled atom-cavity system, as shown in Figure~\ref{fig:setup}.  Cold $^{87}$Rb atoms are trapped at the intersection of two orthogonally aligned standing-wave beams --- a 1030~nm beam focused in the cavity mode with a trap depth of $\approx2.3$~mK and an intracavity standing-wave trap at 785~nm with a trap depth of $\approx30$~$\mu$K \cite{nussmann:2005b, hijlkema:2007}.  Together, these traps create a measured ac-Stark shift of the atomic 5S$_{1/2}\leftrightarrow$~5P$_{3/2}$ transition frequency of approximately +95~MHz.  In addition to providing a second trapping axis, the 785~nm laser is used to stabilize the cavity length to the Stark-shifted D2 $F$$=$$1\leftrightarrow F'$$=$$1$ transition.

The atom-cavity system operates in the intermediate coupling regime with $(g,\kappa,\gamma)/2\pi=(5,6,3)$~MHz, where $g$ denotes the maximum (spatially dependent) atom-cavity coupling constant of the relevant transitions, $\kappa$ is the cavity field decay rate, and $\gamma$ is the atomic polarization decay rate.  Once atoms are loaded into the cavity mode, they are cooled via lin$\bot$lin-polarized laser beams orthogonal to the cavity axis and near resonant with the $F$$=$$2\leftrightarrow F'$$=$$3$ and $F$$=$$1 \leftrightarrow F'$$=$$2$ transitions using a Sisyphus-like cooling mechanism [Fig.~\ref{fig:pulses}(a)] \cite{nussmann:2005b, murr:2006}.  A laser addressing the $F$$=$$1\leftrightarrow F'$$=$$1$ transition is also applied for cavity enhanced cooling and to create photons in the cavity mode.  Photons emitted from the cavity output are coupled into an optical fiber and directed to the photon detection setup.

For high-fidelity entanglement generation, it is important to ensure that exactly one atom is in the cavity.  This is accomplished via two independent techniques.  First, we count the number of trapped atoms by directly imaging the cavity region (Fig.~\ref{fig:setup} and~\ref{fig:pulses}).  A portion of the light scattered by the atoms into free space (perpendicular to the cavity and trapping axes) is collected using an objective lens with a numerical aperture of 0.43, focal length of 25~mm, and a measured resolution of $1.3~\mu$m.  The collected light is focused onto a CCD camera with a total magnification of about 28.  While this technique alone can determine the number of atoms with over 90\% certainty, we further confirm that we have trapped only one atom by analyzing the statistics of the emitted photon stream.  In particular, only if there is exactly one atom in the trap will the cavity output show a perfect photon antibunching signal \cite{hijlkema:2007}.  The combination of these two techniques allows us to discern that a single atom is trapped within the cavity with greater than 99\% fidelity.  

The experimental procedure follows a similar protocol to that used in \cite{wilk:2007b}, but with several substantial differences (Fig.~\ref{fig:pulses}).  First, the trap-induced Stark shift of the atomic energy levels must be taken into account by detuning the laser and cavity frequencies.  Second, the Stark shift has to be stabilized in order to keep the experimental conditions constant, otherwise the fluctuations can lead to unwanted transitions to nearby hyperfine levels of the P$_{3/2}$ manifold.  Moreover, a variable detuning of laser and cavity from the atom decreases the photon generation efficiency.  An additional concern is the random motion of the atom in the dipole trap.  Such motion results from the unidirectional laser pulses employed in the entanglement sequence (discussed below) and shortens the coherence time of atomic superposition states \cite{kuhr:2003}.  In fact, these laser pulses lead to significant heating, expelling the atom from the trap within a few milliseconds.  We find that by embedding each entanglement sequence with an additional cooling interval [Fig.~\ref{fig:pulses}(b)], the atoms remain sufficiently cold to allow for long trapping times and high-fidelity entanglement generation. 

Following this cooling interval, the entanglement protocol starts by optically pumping the atom into the $|F,m_{F}\rangle=|2,0\rangle$ Zeeman sublevel with a measured efficiency greater than $80\%$ [Fig.~\ref{fig:pulses}(b)] \cite{footnote:opticalpump}.  Next, entanglement between the atomic Zeeman state and the polarization of the emitted photon is created by driving a vacuum-stimulated Raman adiabatic passage (vSTIRAP) via a $\pi$-polarized laser pulse addressing the Stark-shifted $F$$=$$2\leftrightarrow$~$F'$$=$$1$ transition and the cavity frequency locked to the $F$$=$$1\leftrightarrow$~$F'$$=$$1$ transition [Fig.~\ref{fig:pulses}(d)] \cite{hennrich:2000}.  With the atom trapped and coupled to the high-finesse optical cavity, the resulting entanglement is inherently deterministic \cite{monroe:2002, kimble:2008}:
\begin{eqnarray}
|\Psi_{\text{AP}}\rangle=\frac{1}{\sqrt{2}}(|1,-1\rangle|\sigma^+\rangle-|1,+1\rangle|\sigma^-\rangle). \label{eq:ap}
\end{eqnarray}
After a user-selected time interval, the atom-photon entanglement is converted into a photon-photon entanglement via a second vSTIRAP step with a $\pi$-polarized $F$$=$$1\leftrightarrow$~$F'$$=$$1$ laser pulse [Fig.~\ref{fig:pulses}(e)].  This maps the atomic state onto the polarization of a second emitted photon, resulting in an entangled photon pair:
\begin{align}
|\Psi_{\text{APP}}\rangle &=|1,0\rangle\otimes|\Psi^-_{\text{PP}}\rangle\nonumber\\
&=\frac{1}{\sqrt{2}}|1,0\rangle\otimes(|\sigma^+\rangle|\sigma^-\rangle-|\sigma^-\rangle|\sigma^+\rangle). \label{eq:pp}
\end{align} 

We characterize our entanglement by measuring a Bell inequality violation of the two emitted photons \cite{bell:1964}.  The form of Bell inequality violated here was first proposed by Clauser, Horne, Shimony, and Holt (CHSH) \cite{clauser:1969}, and is based on the expectation value $E(\alpha,\beta)$ of correlation measurements in different bases:
\begin{align}
E(\alpha,\beta) =&~p_{\downarrow\downarrow}(\alpha,\beta)+p_{\uparrow\uparrow}(\alpha,\beta)\nonumber\\
&-p_{\uparrow\downarrow}(\alpha,\beta)-p_{\downarrow\uparrow}(\alpha,\beta).\label{eq:corr}
\end{align} 
Here, $p_{ij}(\alpha,\beta)$ is the probability to contiguously find photon 1 in state $|i\rangle$ and photon 2 in state $|j\rangle$ following polarization rotations by an amount $\alpha$ and $\beta$, respectively, and $\{\uparrow,\downarrow\}$ represent the two output ports of the polarizing beam splitter.  CHSH show that all local hidden-variable theories must obey the inequality 
\begin{align}
S(\alpha,\alpha';\beta,\beta')\equiv &~|E(\alpha',\beta')-E(\alpha,\beta')|\nonumber\\
&+|E(\alpha',\beta)+E(\alpha,\beta)| \leq 2.\label{eq:ineq}
\end{align} 
This inequality can only be violated via quantum physics.  In particular, our entangled state $|\Psi^-_{\text{PP}}\rangle$ allows for a Bell signal as large as $2\sqrt{2}$.  

\begin{figure}
\includegraphics[width=1\columnwidth,keepaspectratio]{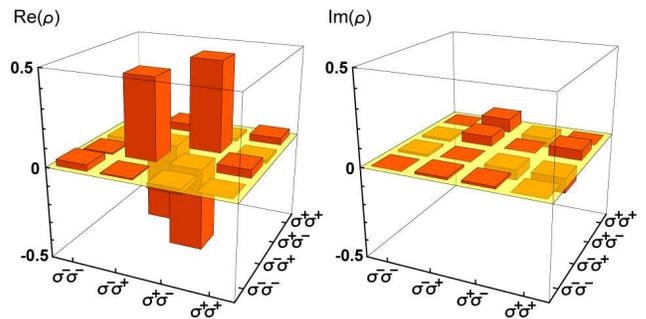}
\caption{\label{fig:dm} 
Real and imaginary parts of the measured two-photon density matrix.  This density matrix represents the $|\Psi^-_{\text{PP}}\rangle$ Bell state of the photons with a fidelity of $F=0.902\pm0.009$.
}
\end{figure}

In our experiment, the two photons are emitted into the same spatial output mode and are probabilistically directed to the two different measurement bases by a 50/50 non-polarizing beam splitter (Fig.~\ref{fig:setup}).  This allows for two simultaneous Bell inequality measurements.  To eliminate possible systematic effects during the course of the experiment, the polarization measurement bases are changed before every atom trapping event via motorized rotation stages.  With the photon pairs measured in a combination of four different polarization bases, we obtain Bell signals of 
\begin{align}
&S(0^{\circ},45^{\circ};22.5^{\circ},-22.5^{\circ})=2.46\pm0.05~\text{and} \nonumber\\
&S(22.5^{\circ},-22.5^{\circ};0^{\circ},45^{\circ})=2.53\pm0.05, \nonumber
\end{align} 
both in clear violation of the classical limit of 2 by more than 9 standard deviations.  In this experiment, the photons are temporally separated by $0.8~\mu$s and the optical path length between the cavity and the photon detectors is 13 meters.  Therefore, the first photon is detected before the generation of the second.  Nevertheless, the fact that the measured correlations violate a Bell inequality can only be explained by quantum entanglement, where the non-classical information between the two photons is temporarily stored in the single trapped atom.  This is similar to experiments with atomic ensembles where the atomic qubit state must be converted to a photon for measurement \cite{jenkins:2007, kimble:2008, yuan:2008}.

The entangled state is additionally characterized via quantum state tomography of the emitted photons.  For this, we follow the procedure outlined in \cite{altepeter:2005} and measure the entangled photons in nine different polarization bases.  The resulting density matrix for the two photons separated by $0.8~\mu$s is shown in Figure~\ref{fig:dm} with an entanglement fidelity of $F=0.902\pm0.009$ with respect to the $|\Psi^-_{\text{PP}}\rangle$ Bell state of the photons (equation~\ref{eq:pp}), clearly above the classical limit of $F=0.5$.  Other calculated measures of entanglement for this state include the concurrence $C=0.81\pm0.03$, entanglement of formation $E_F=0.73\pm0.04$, and logarithmic negativity $E_N=0.867\pm0.014$.  They are all significantly above their classical limit of zero and close to their maximum of 1 for a two-qubit state \cite{plenio:2007}.  From the measured density matrix, we can also infer a Bell signal of $S=2.47\pm0.04$, consistent with the results given above.

\begin{figure}
\includegraphics[width=1\columnwidth,keepaspectratio]{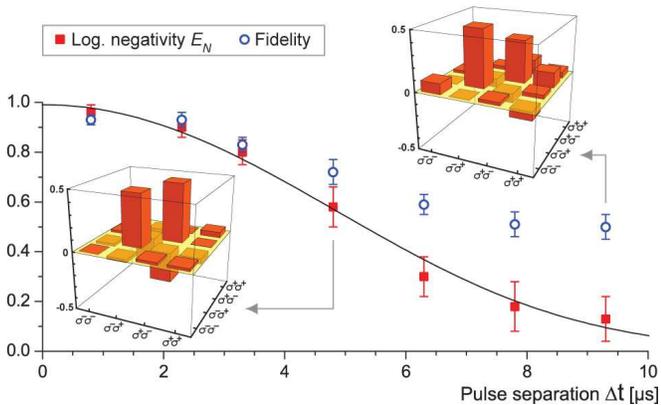}
\caption{\label{fig:coherence} 
Measured fidelity and logarithmic negativity as a function of time between the two vSTIRAP sequences.  Note that this data is independent from that shown in Figure~\ref{fig:dm}.  
The solid line is a Gaussian fit to the negativity $N(\Delta t)$, displayed as $E_N(\Delta t)=\text{log}_2(2N_{\text{o}}e^{-(\Delta t/\tau_e)^2}+1)$, with a resulting entanglement lifetime of $\tau_e=5.7\pm0.2~\mu$s.  The insets show the real parts of the density matrices at differing values of $\Delta t$, (all imaginary parts have a magnitude smaller than 0.14).  The results indicate that the entanglement lifetime is limited mainly by a loss of phase coherence between the $|1,-1\rangle$ and $|1,+1\rangle$ states.
}
\end{figure}

With the atom trapping lifetimes in this experiment of $\approx4.1$ seconds, the separation between the entangling and mapping pulses is currently limited only by the coherence time of the atomic qubit.  This coherence is determined by measuring density matrices as a function of time between the two pulses, $\Delta t$.  We obtain an entanglement lifetime of $\tau_e=5.7\pm0.2~\mu$s (Fig.~\ref{fig:coherence}), limited by phase noise between the two atomic Zeeman states.  This phase sensitivity is evident by the decreasing off-diagonal coherence terms in the density matrix while the diagonal components remain nearly constant (Figures~\ref{fig:dm} and~\ref{fig:coherence}).  This can also be seen from the decay of the fidelity to 50\%, and not 25\% as would be the case for a completely mixed state. 

Our measured entanglement lifetime is comparable to lifetimes observed in atomic ensemble experiments \cite{jenkins:2007, kimble:2008, yuan:2008, simon:2007}.  In our experiment, the limiting mechanisms are magnetic field instabilities ($\sim20$~mG) and a variable differential ac-Stark shift of the atomic superposition states.  The differential Stark shifts are due to motion of the atom together with intensity fluctuations of the cavity stabilization laser ($\sim10\%$) and an uncompensated circular polarization component of the trapping lasers ($\sim2\%$).  With an active stabilization of the magnetic field and optimized laser parameters, this lifetime may be increased to over $100~\mu$s \cite{rosenfeld:2008}.  Additionally, by converting the atomic qubit to clock states, the coherence time of a single atom trapped in a standing-wave can be increased to hundreds of ms \cite{kuhr:2003}.

In addition to the effects mentioned above, the fidelity of the entanglement is further limited by imperfect polarization control in the optical path to the detection setup, dark counts of the photon detectors, and multiple scattered photons during the second pulse.  Indeed, by limiting our photon detection window to include only the first 40\% of the second photon pulse, we observe an increased fidelity of $F=0.932\pm0.014$, albeit with a reduced coincidence rate.  However, with the incorporation of a fast excitation scheme \cite{bochmann:2008} and improved cooling and cavity parameters, many of these effects can be dramatically reduced. 

Finally, the most important aspect for scalable atom-photon networking is the overall success probability.  Here, with a single atom in the cavity, the probability of detecting a two-photon event is about $\approx2.4\times10^{-4}$, as the probability of emitting a single photon into the cavity mode during the entangling pulse and the probability of further emitting a photon during the mapping pulse are each $\approx8.6\%$, and the detection efficiency for a single photon present inside the cavity is $\sim0.2$.  This results in $\approx370$ produced entangled two-photon pairs per second, of which $\approx12$ are detected.  These values are largely limited by the non-optimal atom-cavity coupling due to atomic motion, optical pumping inefficiencies, and photon loss mechanisms, including a 50\% cavity absorption loss due to a mirror defect.  While an atom trapped within an optical cavity can in principle generate photons with unit efficiency \cite{cirac:1997, kimble:2008}, these results still compare well to free-space single atom entanglement experiments with detection probabilities for two subsequent single photons $<5\times10^{-7}$ \cite{matsukevich:2008, rosenfeld:2008}. 

Our entanglement scheme may also be extended to many-photon \cite{schon:2005} and many-atom entanglement protocols \cite{feng:2003, duan:2003, browne:2003}, as well as schemes for quantum teleportation, quantum repeaters \cite{bose:1999}, and a loophole-free Bell inequality violation.  Finally, with the recent completion of a second, independent trapped-atom-cavity system in our group \cite{bochmann:2008}, the demonstration of highly efficient remote-atom entanglement should be possible in the near future.

\begin{acknowledgments}
The authors thank N. Kiesel and A. Ourjoumtsev for useful discussions.  This work was partially supported by the Deutsche Forschungsgemeinschaft (Research Unit 635, Cluster of Excellence MAP) and the European Union (IST project SCALA).  D. L. M. acknowledges support from the Alexander von Humboldt Foundation.
\end{acknowledgments}


\begin{thebibliography}{29}
\expandafter\ifx\csname natexlab\endcsname\relax\def\natexlab#1{#1}\fi
\expandafter\ifx\csname bibnamefont\endcsname\relax
  \def\bibnamefont#1{#1}\fi
\expandafter\ifx\csname bibfnamefont\endcsname\relax
  \def\bibfnamefont#1{#1}\fi
\expandafter\ifx\csname citenamefont\endcsname\relax
  \def\citenamefont#1{#1}\fi
\expandafter\ifx\csname url\endcsname\relax
  \def\url#1{\texttt{#1}}\fi
\expandafter\ifx\csname urlprefix\endcsname\relax\def\urlprefix{URL }\fi
\providecommand{\bibinfo}[2]{#2}
\providecommand{\eprint}[2][]{\url{#2}}

\bibitem[{\citenamefont{Monroe}(2002)}]{monroe:2002}
\bibinfo{author}{\bibfnamefont{C.}~\bibnamefont{Monroe}},
  \bibinfo{journal}{Nature} \textbf{\bibinfo{volume}{416}},
  \bibinfo{pages}{238} (\bibinfo{year}{2002}).

\bibitem[{\citenamefont{Cirac et~al.}(1997)\citenamefont{Cirac, Zoller, Kimble,
  and Mabuchi}}]{cirac:1997}
\bibinfo{author}{\bibfnamefont{J.~I.} \bibnamefont{Cirac}},
  \bibinfo{author}{\bibfnamefont{P.}~\bibnamefont{Zoller}},
  \bibinfo{author}{\bibfnamefont{H.~J.} \bibnamefont{Kimble}},
  \bibnamefont{and} \bibinfo{author}{\bibfnamefont{H.}~\bibnamefont{Mabuchi}},
  \bibinfo{journal}{Phys. Rev. Lett.} \textbf{\bibinfo{volume}{78}},
  \bibinfo{pages}{3221} (\bibinfo{year}{1997}).

\bibitem[{\citenamefont{Kimble}(2008)}]{kimble:2008}
\bibinfo{author}{\bibfnamefont{H.~J.} \bibnamefont{Kimble}},
  \bibinfo{journal}{Nature} \textbf{\bibinfo{volume}{453}},
  \bibinfo{pages}{1023} (\bibinfo{year}{2008}).

\bibitem[{\citenamefont{Blinov et~al.}(2004)\citenamefont{Blinov, Moehring,
  Duan, and Monroe}}]{blinov:2004}
\bibinfo{author}{\bibfnamefont{B.~B.} \bibnamefont{Blinov}},
  \bibinfo{author}{\bibfnamefont{D.~L.} \bibnamefont{Moehring}},
  \bibinfo{author}{\bibfnamefont{L.-M.} \bibnamefont{Duan}}, \bibnamefont{and}
  \bibinfo{author}{\bibfnamefont{C.}~\bibnamefont{Monroe}},
  \bibinfo{journal}{Nature} \textbf{\bibinfo{volume}{428}},
  \bibinfo{pages}{153} (\bibinfo{year}{2004}).

\bibitem[{\citenamefont{Volz et~al.}(2006)\citenamefont{Volz, Weber, Schlenk,
  Rosenfeld, Vrana, Saucke, Kurtsiefer, and Weinfurter}}]{volz:2006}
\bibinfo{author}{\bibfnamefont{J.}~\bibnamefont{Volz}},
  \bibinfo{author}{\bibfnamefont{M.}~\bibnamefont{Weber}},
  \bibinfo{author}{\bibfnamefont{D.}~\bibnamefont{Schlenk}},
  \bibinfo{author}{\bibfnamefont{W.}~\bibnamefont{Rosenfeld}},
  \bibinfo{author}{\bibfnamefont{J.}~\bibnamefont{Vrana}},
  \bibinfo{author}{\bibfnamefont{K.}~\bibnamefont{Saucke}},
  \bibinfo{author}{\bibfnamefont{C.}~\bibnamefont{Kurtsiefer}},
  \bibnamefont{and}
  \bibinfo{author}{\bibfnamefont{H.}~\bibnamefont{Weinfurter}},
  \bibinfo{journal}{Phys. Rev. Lett.} \textbf{\bibinfo{volume}{96}},
  \bibinfo{pages}{030404} (\bibinfo{year}{2006}).

\bibitem[{\citenamefont{Matsukevich et~al.}(2008)\citenamefont{Matsukevich,
  Maunz, Moehring, Olmschenk, and Monroe}}]{matsukevich:2008}
\bibinfo{author}{\bibfnamefont{D.~N.} \bibnamefont{Matsukevich}},
  \bibinfo{author}{\bibfnamefont{P.}~\bibnamefont{Maunz}},
  \bibinfo{author}{\bibfnamefont{D.~L.} \bibnamefont{Moehring}},
  \bibinfo{author}{\bibfnamefont{S.}~\bibnamefont{Olmschenk}},
  \bibnamefont{and} \bibinfo{author}{\bibfnamefont{C.}~\bibnamefont{Monroe}},
  \bibinfo{journal}{Phys. Rev. Lett.} \textbf{\bibinfo{volume}{100}},
  \bibinfo{pages}{150404} (\bibinfo{year}{2008}).

\bibitem[{\citenamefont{Rosenfeld et~al.}(2008)\citenamefont{Rosenfeld, Hocke,
  Henkel, Krug, Volz, Weber, and Weinfurter}}]{rosenfeld:2008}
\bibinfo{author}{\bibfnamefont{W.}~\bibnamefont{Rosenfeld}},
  \bibinfo{author}{\bibfnamefont{F.}~\bibnamefont{Hocke}},
  \bibinfo{author}{\bibfnamefont{F.}~\bibnamefont{Henkel}},
  \bibinfo{author}{\bibfnamefont{M.}~\bibnamefont{Krug}},
  \bibinfo{author}{\bibfnamefont{J.}~\bibnamefont{Volz}},
  \bibinfo{author}{\bibfnamefont{M.}~\bibnamefont{Weber}}, \bibnamefont{and}
  \bibinfo{author}{\bibfnamefont{H.}~\bibnamefont{Weinfurter}},
  \bibinfo{journal}{arXiv:0808.3538v1 [quant-ph]}  (\bibinfo{year}{2008}).

\bibitem[{\citenamefont{Campbell and Benjamin}(2008)}]{campbell:2007}
\bibinfo{author}{\bibfnamefont{E.~T.} \bibnamefont{Campbell}} \bibnamefont{and}
  \bibinfo{author}{\bibfnamefont{S.~C.} \bibnamefont{Benjamin}},
  \bibinfo{journal}{Phys. Rev. Lett.} \textbf{\bibinfo{volume}{101}},
  \bibinfo{eid}{130502} (\bibinfo{year}{2008}).

\bibitem[{\citenamefont{Wilk et~al.}(2007)\citenamefont{Wilk, Webster, Kuhn,
  and Rempe}}]{wilk:2007b}
\bibinfo{author}{\bibfnamefont{T.}~\bibnamefont{Wilk}},
  \bibinfo{author}{\bibfnamefont{S.~C.} \bibnamefont{Webster}},
  \bibinfo{author}{\bibfnamefont{A.}~\bibnamefont{Kuhn}}, \bibnamefont{and}
  \bibinfo{author}{\bibfnamefont{G.}~\bibnamefont{Rempe}},
  \bibinfo{journal}{Science} \textbf{\bibinfo{volume}{317}},
  \bibinfo{pages}{488} (\bibinfo{year}{2007}).

\bibitem[{\citenamefont{Hijlkema et~al.}(2007)\citenamefont{Hijlkema, Weber,
  Specht, Webster, Kuhn, and Rempe}}]{hijlkema:2007}
\bibinfo{author}{\bibfnamefont{M.}~\bibnamefont{Hijlkema}},
  \bibinfo{author}{\bibfnamefont{B.}~\bibnamefont{Weber}},
  \bibinfo{author}{\bibfnamefont{H.~P.} \bibnamefont{Specht}},
  \bibinfo{author}{\bibfnamefont{S.~C.} \bibnamefont{Webster}},
  \bibinfo{author}{\bibfnamefont{A.}~\bibnamefont{Kuhn}}, \bibnamefont{and}
  \bibinfo{author}{\bibfnamefont{G.}~\bibnamefont{Rempe}},
  \bibinfo{journal}{Nature Physics} \textbf{\bibinfo{volume}{3}},
  \bibinfo{pages}{253} (\bibinfo{year}{2007}).

\bibitem[{\citenamefont{Fortier et~al.}(2007)\citenamefont{Fortier, Kim,
  Gibbons, Ahmadi, and Chapman}}]{fortier:2007}
\bibinfo{author}{\bibfnamefont{K.~M.} \bibnamefont{Fortier}},
  \bibinfo{author}{\bibfnamefont{S.~Y.} \bibnamefont{Kim}},
  \bibinfo{author}{\bibfnamefont{M.~J.} \bibnamefont{Gibbons}},
  \bibinfo{author}{\bibfnamefont{P.}~\bibnamefont{Ahmadi}}, \bibnamefont{and}
  \bibinfo{author}{\bibfnamefont{M.~S.} \bibnamefont{Chapman}},
  \bibinfo{journal}{Phys. Rev. Lett.} \textbf{\bibinfo{volume}{98}},
  \bibinfo{eid}{233601} (\bibinfo{year}{2007}).

\bibitem[{\citenamefont{Khudaverdyan et~al.}(2008)\citenamefont{Khudaverdyan,
  Alt, Dotsenko, Kampschulte, Lenhard, Rauschenbeutel, Reick, Sch\"{o}rner,
  Widera, and Meschede}}]{khudaverdyan:2008}
\bibinfo{author}{\bibfnamefont{M.}~\bibnamefont{Khudaverdyan}},
  \bibinfo{author}{\bibfnamefont{W.}~\bibnamefont{Alt}},
  \bibinfo{author}{\bibfnamefont{I.}~\bibnamefont{Dotsenko}},
  \bibinfo{author}{\bibfnamefont{T.}~\bibnamefont{Kampschulte}},
  \bibinfo{author}{\bibfnamefont{K.}~\bibnamefont{Lenhard}},
  \bibinfo{author}{\bibfnamefont{A.}~\bibnamefont{Rauschenbeutel}},
  \bibinfo{author}{\bibfnamefont{S.}~\bibnamefont{Reick}},
  \bibinfo{author}{\bibfnamefont{K.}~\bibnamefont{Sch\"{o}rner}},
  \bibinfo{author}{\bibfnamefont{A.}~\bibnamefont{Widera}}, \bibnamefont{and}
  \bibinfo{author}{\bibfnamefont{D.}~\bibnamefont{Meschede}},
  \bibinfo{journal}{New Journal of Physics} \textbf{\bibinfo{volume}{10}},
  \bibinfo{pages}{073023} (\bibinfo{year}{2008}).

\bibitem[{\citenamefont{Bell}(1964)}]{bell:1964}
\bibinfo{author}{\bibfnamefont{J.~S.} \bibnamefont{Bell}},
  \bibinfo{journal}{Physics (Long Island City, N.Y.)}
  \textbf{\bibinfo{volume}{1}}, \bibinfo{pages}{195} (\bibinfo{year}{1964}).

\bibitem[{\citenamefont{Nu{\ss}mann et~al.}(2005)\citenamefont{Nu{\ss}mann,
  Murr, Hijlkema, Weber, Kuhn, and Rempe}}]{nussmann:2005b}
\bibinfo{author}{\bibfnamefont{S.}~\bibnamefont{Nu{\ss}mann}},
  \bibinfo{author}{\bibfnamefont{K.}~\bibnamefont{Murr}},
  \bibinfo{author}{\bibfnamefont{M.}~\bibnamefont{Hijlkema}},
  \bibinfo{author}{\bibfnamefont{B.}~\bibnamefont{Weber}},
  \bibinfo{author}{\bibfnamefont{A.}~\bibnamefont{Kuhn}}, \bibnamefont{and}
  \bibinfo{author}{\bibfnamefont{G.}~\bibnamefont{Rempe}},
  \bibinfo{journal}{Nature Physics} \textbf{\bibinfo{volume}{1}},
  \bibinfo{pages}{122} (\bibinfo{year}{2005}).

\bibitem[{\citenamefont{Murr et~al.}(2006)\citenamefont{Murr, Nu{\ss}mann,
  Puppe, Hijlkema, Weber, Webster, Kuhn, and Rempe}}]{murr:2006}
\bibinfo{author}{\bibfnamefont{K.}~\bibnamefont{Murr}},
  \bibinfo{author}{\bibfnamefont{S.}~\bibnamefont{Nu{\ss}mann}},
  \bibinfo{author}{\bibfnamefont{T.}~\bibnamefont{Puppe}},
  \bibinfo{author}{\bibfnamefont{M.}~\bibnamefont{Hijlkema}},
  \bibinfo{author}{\bibfnamefont{B.}~\bibnamefont{Weber}},
  \bibinfo{author}{\bibfnamefont{S.~C.} \bibnamefont{Webster}},
  \bibinfo{author}{\bibfnamefont{A.}~\bibnamefont{Kuhn}}, \bibnamefont{and}
  \bibinfo{author}{\bibfnamefont{G.}~\bibnamefont{Rempe}},
  \bibinfo{journal}{Phys. Rev. A} \textbf{\bibinfo{volume}{73}},
  \bibinfo{eid}{063415} (\bibinfo{year}{2006}).

\bibitem[{\citenamefont{Kuhr et~al.}(2003)\citenamefont{Kuhr, Alt, Schrader,
  Dotsenko, Miroshnychenko, Rosenfeld, Khudaverdyan, Gomer, Rauschenbeutel, and
  Meschede}}]{kuhr:2003}
\bibinfo{author}{\bibfnamefont{S.}~\bibnamefont{Kuhr}},
  \bibinfo{author}{\bibfnamefont{W.}~\bibnamefont{Alt}},
  \bibinfo{author}{\bibfnamefont{D.}~\bibnamefont{Schrader}},
  \bibinfo{author}{\bibfnamefont{I.}~\bibnamefont{Dotsenko}},
  \bibinfo{author}{\bibfnamefont{Y.}~\bibnamefont{Miroshnychenko}},
  \bibinfo{author}{\bibfnamefont{W.}~\bibnamefont{Rosenfeld}},
  \bibinfo{author}{\bibfnamefont{M.}~\bibnamefont{Khudaverdyan}},
  \bibinfo{author}{\bibfnamefont{V.}~\bibnamefont{Gomer}},
  \bibinfo{author}{\bibfnamefont{A.}~\bibnamefont{Rauschenbeutel}},
  \bibnamefont{and} \bibinfo{author}{\bibfnamefont{D.}~\bibnamefont{Meschede}},
  \bibinfo{journal}{Phys. Rev. Lett.} \textbf{\bibinfo{volume}{91}},
  \bibinfo{pages}{213002} (\bibinfo{year}{2003}).

\bibitem{footnote:opticalpump}
Note that optical pumping to the wrong state will not reduce the measured entanglement fidelity 
as it cannot result in a two-photon entanglement event \cite{wilk:2007b}.
  
\bibitem[{\citenamefont{Hennrich et~al.}(2000)\citenamefont{Hennrich, Legero,
  Kuhn, and Rempe}}]{hennrich:2000}
\bibinfo{author}{\bibfnamefont{M.}~\bibnamefont{Hennrich}},
  \bibinfo{author}{\bibfnamefont{T.}~\bibnamefont{Legero}},
  \bibinfo{author}{\bibfnamefont{A.}~\bibnamefont{Kuhn}}, \bibnamefont{and}
  \bibinfo{author}{\bibfnamefont{G.}~\bibnamefont{Rempe}},
  \bibinfo{journal}{Phys. Rev. Lett.} \textbf{\bibinfo{volume}{85}},
  \bibinfo{pages}{4872} (\bibinfo{year}{2000}).

\bibitem[{\citenamefont{Clauser et~al.}(1969)\citenamefont{Clauser, Horne,
  Shimony, and Holt}}]{clauser:1969}
\bibinfo{author}{\bibfnamefont{J.~F.} \bibnamefont{Clauser}},
  \bibinfo{author}{\bibfnamefont{M.~A.} \bibnamefont{Horne}},
  \bibinfo{author}{\bibfnamefont{A.}~\bibnamefont{Shimony}}, \bibnamefont{and}
  \bibinfo{author}{\bibfnamefont{R.~A.} \bibnamefont{Holt}},
  \bibinfo{journal}{Phys. Rev. Lett.} \textbf{\bibinfo{volume}{23}},
  \bibinfo{pages}{880} (\bibinfo{year}{1969}).

\bibitem[{\citenamefont{Jenkins et~al.}(2007)\citenamefont{Jenkins,
  Matsukevich, Chanelière, Lan, Kennedy, and Kuzmich}}]{jenkins:2007}
\bibinfo{author}{\bibfnamefont{S.~D.} \bibnamefont{Jenkins}},
  \bibinfo{author}{\bibfnamefont{D.~N.} \bibnamefont{Matsukevich}},
  \bibinfo{author}{\bibfnamefont{T.}~\bibnamefont{Chanelière}},
  \bibinfo{author}{\bibfnamefont{S.-Y.} \bibnamefont{Lan}},
  \bibinfo{author}{\bibfnamefont{T.~A.~B.} \bibnamefont{Kennedy}},
  \bibnamefont{and} \bibinfo{author}{\bibfnamefont{A.}~\bibnamefont{Kuzmich}},
  \bibinfo{journal}{J. Opt. Soc. Am. B} \textbf{\bibinfo{volume}{24}},
  \bibinfo{pages}{316} (\bibinfo{year}{2007}).

\bibitem[{\citenamefont{Yuan et~al.}(2008)\citenamefont{Yuan, Chen, Zhao, Chen,
  Schmiedmayer, and Pan}}]{yuan:2008}
\bibinfo{author}{\bibfnamefont{Z.-S.} \bibnamefont{Yuan}},
  \bibinfo{author}{\bibfnamefont{Y.-A.} \bibnamefont{Chen}},
  \bibinfo{author}{\bibfnamefont{B.}~\bibnamefont{Zhao}},
  \bibinfo{author}{\bibfnamefont{S.}~\bibnamefont{Chen}},
  \bibinfo{author}{\bibfnamefont{J.}~\bibnamefont{Schmiedmayer}},
  \bibnamefont{and} \bibinfo{author}{\bibfnamefont{J.-W.} \bibnamefont{Pan}},
  \bibinfo{journal}{Nature} \textbf{\bibinfo{volume}{454}},
  \bibinfo{pages}{1098} (\bibinfo{year}{2008}).

\bibitem[{\citenamefont{Altepeter et~al.}(2005)\citenamefont{Altepeter,
  Jeffrey, and Kwiat}}]{altepeter:2005}
\bibinfo{author}{\bibfnamefont{J.~B.} \bibnamefont{Altepeter}},
  \bibinfo{author}{\bibfnamefont{E.~R.} \bibnamefont{Jeffrey}},
  \bibnamefont{and} \bibinfo{author}{\bibfnamefont{P.~G.} \bibnamefont{Kwiat}},
  \bibinfo{journal}{Adv. At. Mol. Opt. Phys.} \textbf{\bibinfo{volume}{52}},
  \bibinfo{pages}{105} (\bibinfo{year}{2005}).

\bibitem[{\citenamefont{Plenio and Virmani}(2007)}]{plenio:2007}
\bibinfo{author}{\bibfnamefont{M.~B.} \bibnamefont{Plenio}} \bibnamefont{and}
  \bibinfo{author}{\bibfnamefont{S.}~\bibnamefont{Virmani}},
  \bibinfo{journal}{Quant. Inf. Comp.} \textbf{\bibinfo{volume}{7}},
  \bibinfo{pages}{1} (\bibinfo{year}{2007}).

\bibitem[{\citenamefont{Simon et~al.}(2007)\citenamefont{Simon, Tanji, Ghosh,
  and Vuleti\'{c}}}]{simon:2007}
\bibinfo{author}{\bibfnamefont{J.}~\bibnamefont{Simon}},
  \bibinfo{author}{\bibfnamefont{H.}~\bibnamefont{Tanji}},
  \bibinfo{author}{\bibfnamefont{S.}~\bibnamefont{Ghosh}}, \bibnamefont{and}
  \bibinfo{author}{\bibfnamefont{V.}~\bibnamefont{Vuleti\'{c}}},
  \bibinfo{journal}{Nature Physics} \textbf{\bibinfo{volume}{3}},
  \bibinfo{pages}{1745} (\bibinfo{year}{2007}).

\bibitem[{\citenamefont{Bochmann et~al.}(2008)\citenamefont{Bochmann,
  M{\"{u}}cke, Langfahl-Klabes, Erbel, Weber, Specht, Moehring, and
  Rempe}}]{bochmann:2008}
\bibinfo{author}{\bibfnamefont{J.}~\bibnamefont{Bochmann}},
  \bibinfo{author}{\bibfnamefont{M.}~\bibnamefont{M{\"{u}}cke}},
  \bibinfo{author}{\bibfnamefont{G.}~\bibnamefont{Langfahl-Klabes}},
  \bibinfo{author}{\bibfnamefont{C.}~\bibnamefont{Erbel}},
  \bibinfo{author}{\bibfnamefont{B.}~\bibnamefont{Weber}},
  \bibinfo{author}{\bibfnamefont{H.~P.} \bibnamefont{Specht}},
  \bibinfo{author}{\bibfnamefont{D.~L.} \bibnamefont{Moehring}},
  \bibnamefont{and} \bibinfo{author}{\bibfnamefont{G.}~\bibnamefont{Rempe}},
  \bibinfo{journal}{(Phys. Rev. Lett. in press) arXiv:0806.2600v1
  [quant-ph]}  (\bibinfo{year}{2008}).

\bibitem[{\citenamefont{Sch{\"{o}}n et~al.}(2005)\citenamefont{Sch{\"{o}}n,
  Solano, Verstraete, Cirac, and Wolf}}]{schon:2005}
\bibinfo{author}{\bibfnamefont{C.}~\bibnamefont{Sch{\"{o}}n}},
  \bibinfo{author}{\bibfnamefont{E.}~\bibnamefont{Solano}},
  \bibinfo{author}{\bibfnamefont{F.}~\bibnamefont{Verstraete}},
  \bibinfo{author}{\bibfnamefont{J.~I.} \bibnamefont{Cirac}}, \bibnamefont{and}
  \bibinfo{author}{\bibfnamefont{M.~M.} \bibnamefont{Wolf}},
  \bibinfo{journal}{Phys. Rev. Lett.} \textbf{\bibinfo{volume}{95}},
  \bibinfo{eid}{110503} (\bibinfo{year}{2005}).

\bibitem[{\citenamefont{Feng et~al.}(2003)\citenamefont{Feng, Zhang, Li, Gong,
  and Xu}}]{feng:2003}
\bibinfo{author}{\bibfnamefont{X.-L.} \bibnamefont{Feng}},
  \bibinfo{author}{\bibfnamefont{Z.-M.} \bibnamefont{Zhang}},
  \bibinfo{author}{\bibfnamefont{X.-D.} \bibnamefont{Li}},
  \bibinfo{author}{\bibfnamefont{S.-Q.} \bibnamefont{Gong}}, \bibnamefont{and}
  \bibinfo{author}{\bibfnamefont{Z.-Z.} \bibnamefont{Xu}},
  \bibinfo{journal}{Phys. Rev. Lett.} \textbf{\bibinfo{volume}{90}},
  \bibinfo{pages}{217902} (\bibinfo{year}{2003}).

\bibitem[{\citenamefont{Duan and Kimble}(2003)}]{duan:2003}
\bibinfo{author}{\bibfnamefont{L.-M.} \bibnamefont{Duan}} \bibnamefont{and}
  \bibinfo{author}{\bibfnamefont{H.~J.}~\bibnamefont{Kimble}},
  \bibinfo{journal}{Phys. Rev. Lett.} \textbf{\bibinfo{volume}{90}},
  \bibinfo{pages}{253601} (\bibinfo{year}{2003}).

\bibitem[{\citenamefont{Browne et~al.}(2003)\citenamefont{Browne, Plenio, and
  Huelga}}]{browne:2003}
\bibinfo{author}{\bibfnamefont{D.~E.} \bibnamefont{Browne}},
  \bibinfo{author}{\bibfnamefont{M.~B.} \bibnamefont{Plenio}},
  \bibnamefont{and} \bibinfo{author}{\bibfnamefont{S.~F.}
  \bibnamefont{Huelga}}, \bibinfo{journal}{Phys. Rev. Lett.}
  \textbf{\bibinfo{volume}{91}}, \bibinfo{pages}{067901}
  (\bibinfo{year}{2003}).

\bibitem[{\citenamefont{Bose et~al.}(1999)\citenamefont{Bose, Knight, Plenio,
  and Vedral}}]{bose:1999}
\bibinfo{author}{\bibfnamefont{S.}~\bibnamefont{Bose}},
  \bibinfo{author}{\bibfnamefont{P.~L.} \bibnamefont{Knight}},
  \bibinfo{author}{\bibfnamefont{M.~B.} \bibnamefont{Plenio}},
  \bibnamefont{and} \bibinfo{author}{\bibfnamefont{V.}~\bibnamefont{Vedral}},
  \bibinfo{journal}{Phys. Rev. Lett.} \textbf{\bibinfo{volume}{83}},
  \bibinfo{pages}{5158} (\bibinfo{year}{1999}).

\end{thebibliography}

\end{document}